\documentclass[10pt]{article}
\usepackage[hidelinks]{hyperref}

\newcommand{\orcid}[1]{\href{https://orcid.org/#1}{}} 

\title{A Serverless Distributed Ledger for Enterprises}

\newif\ifblind
\blindfalse
\ifblind
\author{}
\else
\author{Johannes Sedlmeir, Tim Wagner, Emil Djerekarov, \\Ryan Green, Johannes Klepsch, Shruthi Rao}

\usepackage{hyperref}
\usepackage{graphicx}
\usepackage{comment}
\usepackage{xcolor}
\usepackage[nolist]{acronym}
\usepackage{eurosym}
\usepackage{booktabs}
\usepackage{float}
\usepackage{amsmath}
\usepackage{boxedminipage}
\usepackage{xurl}
\usepackage{enumitem}

\usepackage{xcolor}

\newif\ifdraft
\drafttrue
\ifdraft
\newcommand{\jsnote}[1]{ {\textcolor{blue} { ***Johannes S.: #1 }}}
\newcommand{\twnote}[1]{ {\textcolor{orange} { ***Tim: #1 }}}
\newcommand{\ednote}[1]{ {\textcolor{green} { ***Emil: #1 }}}
\newcommand{\jknote}[1]{ {\textcolor{red} { ***Johannes K.: #1 }}}
\else
\newcommand{\jsnote}[1]{}
\newcommand{\twnote}[1]{}
\newcommand{\ednote}[1]{}
\newcommand{\jknote}[1]{}
\fi

\usepackage{array}
\usepackage{booktabs}
\newcolumntype{P}[1]{>{\centering\arraybackslash}p{#1}}
\newcolumntype{L}[1]{>{\raggedright\arraybackslash}p{#1}}
\newcolumntype{C}[1]{>{\centering\arraybackslash}p{#1}}

\usepackage{pifont}

\begin{acronym}
\acro{ACL}{access control list}
\acro{API}{application programming interface}
\acro{AWS}{Amazon Web Services}
\acro{AZ}{availability zone}
\acro{BFT}{Byzantine fault tolerant}
\acro{CCPA}{California Consumer Privacy Act}
\acro{CDK}{cloud development kit}
\acro{CFT}{crash fault tolerant}
\acro{CSP}{cloud service provider}
\acro{DDOS}{distributed-denial-of-service}
\acro{DL}{distributed ledger}
\acro{DLPS}{distributed ledger performance scan}
\acro{DLT}{distributed ledger technology}
\acroplural{DLT}[DLTs]{distributed ledger technologies}
\acro{EC2}{elastic compute cloud}
\acro{GCP}{Google Cloud Platform}
\acro{GDPR}{general data protection regulation}
\acro{IAM}{identity and access management}
\acro{OLTP}{online transaction processing}
\acro{PII}{personally identifiable information}
\acro{PBFT}{practical BFT}
\acro{PoW}{proof of work}
\acro{PoS}{proof of stake}
\acro{PKI}{public key infrastructure}
\acro{SaaS}{software as a service}
\acro{SLA}{service level agreement}
\acro{SQS}{simple queue service}
\acro{VM}{virtual machine}
\acro{VPC}{virtual private cloud}
\acro{ZKP}{zero-knowledge proof}
\end{acronym}

\usepackage{tabularx}  

\begin{document}

\clearpage

\maketitle

\noindent This paper has been accepted at the 55th Hawaii International Conference on System Sciences (HICSS) and will be published in January 2022. \\\noindent

\begin{abstract}
Enterprises have been attracted by the capability of blockchains to provide a single source of truth for workloads that span companies, geographies, and clouds while retaining the independence of each party’s IT operations. However, so far production applications have remained rare, stymied by technical limitations of existing blockchain technologies and challenges with their integration into enterprises' IT systems. 
In this paper, we collect enterprises' requirements on distributed ledgers for data sharing and integration from a technical perspective, argue that they are not sufficiently addressed by available blockchain frameworks, and propose a novel distributed ledger design that is ``serverless'', i.e., built on cloud-native resources. We evaluate its qualitative and quantitative properties and give evidence that enterprises already heavily reliant on cloud service providers would consider such an approach acceptable, particularly if it offers ease of deployment, low transactional cost structure, and a combination of latency and scalability aligned with real-time IT application needs.
\end{abstract}

\section{Introduction}
\label{sec:introduction}
Data, particularly transactional data housed in various flavors of databases, powers the vast majority of modern IT applications. Historically, the bulk of that data was produced and consumed by the owner of the data. However, the growing complexity of supply and logistics chains, the ``consumerization'' of IT bringing ever-higher expectations for real-time information and automated decision making, and the trend towards simplified \ac{SaaS} deployments are all causing data to migrate outside a company's four walls. Classic mechanisms to provide trustworthy, high-fidelity data representation and query results, such as centralized databases offering ACID transactions and SQL query languages, fail when a considerable fraction of data resides elsewhere, accessible only through batch files or by polling third-party APIs, making it potentially inconsistent, incomplete, and out of date. Blockchain technologies appeared to offer a compelling solution to this challenge: A technology that could simultaneously erect a single source of truth in the form of a distributed ledger capable of spanning companies, clouds, and geographical boundaries, while still preserving each individual participant's control over its own technology stack, including deployment, authentication, security, and compliance needs.

Distributed ledgers used to create distributed, multi-party databases with ACID semantics have a lengthy research history. As far back as the 1980s, researchers investigated \ac{CFT} and \ac{BFT} state machine replication in order to achieve reliable distributed systems in the presence of failures or adversaries~\cite{lamport1982byzantine}. The security of these systems was based on an election mechanism in a \emph{permissioned} setting (two- or three-phase commit), where the identities of all participants or at least their total number was known. Although the first, merely crash-fault tolerant solutions such as Paxos were soon improved, e.g., through Byzantine-fault tolerant protocols such as PBFT~\cite{Castro:1999:pbft}, direct adoption of \acp{DLT} by enterprises remained rare until recently, although \emph{indirect} usage in the form of public cloud databases that make use of permissioned consensus and Paxos variants became commonplace as cloud adoption has grown~\cite{ailijiang2016consensus}. 

The original Bitcoin whitepaper~\cite{Nakamoto:2008:Bitcoin} popularized a \emph{permissionless} \ac{DLT} combined with Sybil attack prevention for the purposes of value storage and transfer that has come to be known as a cryptocurrency. Ethereum expanded on the simple ``value transfer" interpreter in Bitcoin with a Turing complete computational engine or Smart Contract platform~\cite{Buterin:2014:Ethereum}. Ethereum garnered attention outside the cryptocurrency and speculative financial market communities through its self-marketing as the ``world computer". Enterprises and private sector consortia eager for solutions to the inconsistencies, omissions, and high manual reconciliation costs of data silos looked to Ethereum and its variants as a possible solution. At the same time, information systems researchers were attracted by applications of \ac{DLT} that promised businesses considerable improvements in terms of interoperability, traceability, provenance, distributed control, accountability, and transparency~\cite{beck2018governance} by providing a neutral digital infrastructure for cross-organizational processes~\cite{fridgen2018cross}. Unfortunately, the duplication of computation and storage on every node in the network~\cite{luu2015demystifying} as well as the need for economic incentives imply low throughput, high latency, and significant transaction costs~\cite{kannengiesser2020trade} and thus make the public permissionless blockchains a non-starter for the vast majority of enterprise use cases, even ignoring potential concerns about exposing their data to the world~\cite{zhang2019security}.

Consequently, enterprises have generally found more success with \emph{permissioned} blockchain networks in various sectors, e.g., in improving data exchange and traceability in automotive supply chains~\cite{miehle2019partchain}. Popular open-source implementations of permissioned \acp{DLT} include private Ethereum networks such as Quorum, and Hyperledger Fabric~\cite{androulaki2018hyperledger}. Permissioned blockchains provide many advantages over permissionless blockchains for enterprises, including higher performance, predictable costs and the support of data confidentiality features ``off-the-shelf''.
Despite these relative advantages, the performance of permissioned blockchains still remains orders of magnitude lower than ``centralized" database technologies~\cite{barr_2019}, and -- as we will argue in this paper -- the costs and complexities associated with setting up and maintaining \acp{DLT} for enterprises are significant. 

We posit that many of the limitations regarding performance, complexity, and cost in existing enterprise distributed ledger implementations are driven by their reliance on a server-based deployment model and suggest an intriguing alternative: a distributed ledger in which each node is built using ``serverless'' infrastructure~\cite{castro2019rise}, thus benefiting from the economic and scaling advantages of massive multi-tenanted implementations that expose inter-machine parallelism opportunities unavailable to prior techniques. Our approach offers the performance and ``form fit'' of a cloud-based SQL or NoSQL database approach while retaining the decentralized aspect of a permissioned blockchain in the form of segregated accounts containing individually owned resources, in exchange for giving up the ability to run nodes outside of a public cloud setting.

The remainder of this paper is structured as follows: In section~\ref{sec:background} we briefly review serverless architectures and survey related work. Section~\ref{sec:requirements} derives common enterprise requirements for blockchains used for data integration purposes. In section~\ref{sec:serverless}, we present the main components and characteristics of our serverless blockchain architecture. We then evaluate our implementation from a qualitative and quantitative perspective in sections~\ref{sec:qualitative_analysis} and~\ref{sec:quantitative_analysis}. We summarize our observations and avenues for further research in section~\ref{sec:conclusion}.

\section{Background}
\label{sec:background}
\subsection{Serverless computing}

Surveys of serverless offerings and research describe cloud-based compute, storage, queuing, \ac{API} hosting, and workflow (choreography) services that offer access to effectively unbounded storage and compute power coupled with a pay-per-call cost structure and latency on the order of 8-10\,ms~\cite{jonas2019cloud}. The massively multi-tenanted nature of these services provides an alternative to blockchain algorithms constrained to using a single server per node: Effectively, such a system can ``dispatch'' thousands of virtual machines in single digit milliseconds, each one verifying or applying an individual transaction within a block. Coupled to the massively parallel front ends of NoSQL databases and blob storage, end-to-end processing and storage parallelism enables individual blockchain nodes to escape the confines of vertical scaling and the prohibitive cost dynamics of scaling each node to peak needs at all times. Reconstructing consensus out of these building blocks exposes multi-machine parallelism opportunities not available in extant blockchain approaches, particularly as conventional consensus algorithms also consider the machine on which they run to be an atomic unit of trust and network identity.

The term ``serverless'' has entered the lexicon to denote services and architectures that rely on fully managed cloud services~\cite{schleier-smith2021serverless}. Compared to older application construction methodologies in which companies rent servers from \acp{CSP} such as \ac{AWS}, Azure, or Google, serverless architectures rely on the use of services that hide the presence of servers beyond an abstraction layer~\cite{castro2019rise}. \ac{AWS}~Lambda, a serverless cloud computing service introduced in November~2014, initiated much of the current interest in the category. Lambda works by multi-tenanting both at the fleet and the individual machine level, placing hypervisors around each workload. 

Computations are invoked by HTTPS requests and routed to a (possibly preexisting) container by a low single-digit millisecond bin packing router that is tenant aware.
Cloud function services from other \acp{CSP} work similarly. Serverless \ac{CSP} services, and the applications constructed using them, are typically differentiated along several dimensions of interest to our analysis:
\begin{itemize}[leftmargin=12pt]
    \item \emph{Intrinsically fault tolerant:} The ``gold standard" for a highly available (99.9 -- 99.99\,\% uptime) system is 3-way redundancy across spatially isolated data centers -- what \ac{AWS} refers to as \acp{AZ}. 
    Serverless offerings hence build fault tolerance into their implementation, so the service \emph{as delivered to the application} includes redundancy by design. By contrast, each participant in a conventional blockchain would need to own and operate at least three nodes themselves, just to ensure an equivalent availability outcome.
    \item \emph{Scale-per-request:} With a conventional architecture, scaling up the operational capability of a system requires either vertical or horizontal scaling techniques; i.e., either ``rent a bigger box" or ``rent more boxes". Serverless services manage that scaling behind the scenes, typically relying on massively multi-tenanted fleets, which provides the illusion of essentially limitless scaling driven exclusively through making requests to the service's \ac{API}.
    \item \emph{Pay-per-request:} Serverless offerings typically charge on a per-request basis (rather than a time-based rental fee), and thus unlike infrastructure-based architectures they ``turn off" completely, generating no costs when not in use~\cite{castro2019rise}. Given that typical enterprise fleets only achieve around 18\,\% utilization~\cite{451ResearchUtilization}, this can represent a significant improvement in costs and also energy consumption.
\end{itemize}

\subsection{Related work}
Researchers have already started to analyze the tradeoffs and challenges that come with blockchain adoption from a technical~\cite{kannengiesser2020trade} and organizational perspective~\cite{zavolokina2020management}. Moreover, work like~\cite{lacity2018addressing} focused on structuring standardization, performance, and regulatory requirements and developed strategies to address associated challenges. Recently, publications have analyzed the business-related challenges in specific application areas, such as supply chains~\cite{hastig2020blockchain}. \cite{chatterjee2019challenges} discusses challenges of \ac{DLT} adoption from the perspective of enterprises from a review of literature based on a weighted average score. However, to our knowledge, so far no large-scale study involving enterprises has been conducted to determine enterprises' requirements on integrating distributed ledger technologies in their IT infrastructure.

To locate additional related work that aims to propose a serverless blockchain design, we applied the search term ``serverless AND (blockchain OR distributed ledger)'' in the \href{https://arxiv.org/}{arXiv}, \href{https://dl.acm.org/}{ACM digital library},  \href{https://scholar.google.com/}{Google scholar}, and \href{https://ieeexplore.ieee.org/Xplore/home.jsp}{IEEE Xplore} databases. We found that~\cite{oh2019serverless} suggested an architecture in which some computational tasks for the Hyperledger Sawtooth blockchain can be shifted to \ac{AWS}~Lambda. Beyond this, \cite{ghaemi2020chainfaas}~uses a serverless approach in another sense, namely shifting computational work in the context of a blockchain to a user's personal devices. Finally, \cite{kaplunovich2019scalability} use \ac{AWS}~Lambda for the client side, sending requests to a Fabric network for a performance analysis. Recently, a cryptographically verifiable SQL Database for Azure has been proposed~\cite{antonopoulos2021azure} that has some similarities with the storage layer of the serverless blockchain that we propose; however, this service is restricted to a single account and hence does not explore consensus-related topics that could support the synchronization of data across multiple accounts. Consequently, we found no work that suggests an architecture for or creates a fully functional distributed ledger based on serverless components.
\section{Business requirements for blockchains}
\label{sec:requirements}
To collect requirements for enterprise blockchains for data sharing and integration from practitioners, we interviewed~1,092 companies having at least some prior public cloud experience in person from~2017 through~2019. They spanned enterprises, SMBs, and startups and represent verticals such as automotive, financial services, consumer packaged goods, food \& beverage, travel \& hospitality, media \& entertainment, agriculture, IT, telecom and semiconductors, and public sector. Over 95\,\% purchased services from \ac{AWS} and over 91\,\% spent at least \$50,000 per month on cloud services. 98\,\% of interviewed companies were for-profit and nearly two thirds were enterprises. All interviews were 60~minutes or more in duration and included both structured feedback and free-form inquiry regarding data  sharing and application construction requirements, intended use cases for blockchain or ledger-like offerings, and -- where applicable -- reasons for adopting or abandoning blockchain technology. We did not select interviewees specifically for success or failure of \ac{DLT} projects, but all of our interviewees had expressed interest in, or were actively involved with, a \ac{DLT} project.

\begin{table}[!htb]
    \centering
    \resizebox{\linewidth}{!}{
    \begin{tabular}{p{3.2cm}|p{10.9cm}}
    \textbf{Capability} & \textbf{Requirements}\\\midrule 
    Decentralization & Each participant must be able to maintain a legally and operationally independent copy of all data and metadata without reliance on another company's IT organization.
    \\\midrule
    (Multi-)Cloud \mbox{Deployment} & \ac{DLT}s
    used for data integration purposes must be deployable to public clouds in order to integrate with existing IT security and operations. Each node must also be able to make an independent decision with respect to the choice of \ac{CSP}, enabling participants to achieve low-latency interconnect with other resources and services in that \ac{CSP}.\\\midrule
    \mbox{Elastic Scaling} \mbox{on Demand} & The \ac{DLT} must support flexible and effectively
    instantaneous scalability to accommodate enterprise IT workloads, which may vary unpredictably. 
    A cost structure that scales linearly (versus being scaled perpetually to peak capacity) is a positive differentiator. \\\midrule
    Unlimited \break Storage & Enterprises expect \ac{DLT}s used as data storage and integration
    solutions to operate without limits on any form of storage (file, blob, database size, etc.). \\\midrule
    \raggedright{Fault Tolerance and High Availability} & 99.99\,\% availability is the standard for enterprise contracts, with mission critical and financial systems often requiring 99.999\,\% uptime. Having the \ac{DLT} provide this capability intrinsically with no additional cost, deployment complexity, or maintenance on the part of the user is a positive differentiator. \\\midrule
    \mbox{Ease of}\break Deployment &
    Conventional blockchain deployments often demand non-trivial
    staffing to configure, deploy, and maintain.
    Schema-driven definition (similar to conventional database
    tables), \ac{SaaS}-based delivery, and limiting
    manual labor required for networking, operating system, virtual machine, security, or availability configuration are positive differentiators.\\\midrule
    \hbox{Low Latency} and \hbox{Fast Finality} & Many \ac{OLTP} tasks in enterprise applications have near real-time data processing expectations, requiring fast (sub-second) confirmation of transactions.\\\midrule
    Energy Efficiency & The \ac{DLT}'s effective utilization, and its reliance on the power grid, must be in line with typical corporate applications. Improved utilization relative to state of the practice is a differentiator. \\\midrule
    \raggedright{Access Control and Data Governance} & Nearly all enterprises require the ability to scope ledger and world state updates to a subset of participants on a per-transaction basis (``private transactions"), for reasons ranging from
    business confidentiality, to material disclosure laws,
    to data protection regulations.
    \\\midrule
    \end{tabular}
    }
    \caption{Business requirements for \ac{DLT}s used for enterprise data sharing and integration applications.}
    \label{tab:requirements}
\end{table}

The most frequently cited reason for adopting (or intending to adopt, as was more often the case) blockchain technology was what we termed ``dispersed data'' problems: Internal data that spanned departments and/or multi-company workflows that spanned business partners, such as suppliers or logistics. Frequently, this data also had to traverse at least one other divide: multiple geographies, multiple providers (\ac{AWS}, Azure, Snowflake, and Databricks were the most frequently cited), or needed to straddle an on-premise/public cloud connection. Interviewees often chose terms such as, "single source of truth", "shared system of record", "breaking down data silos", "connecting data", or "multi-party solutions" to express their desired end states and their reason for considering blockchain as a solution. Highly correlated requirements included privacy and security concerns, with interviewees often stressing that some form of access controls were mandatory to enable them to ``keep control of their data'', and the public blockchains were thus often a non starter as a place to store actual business data. Secondary concerns included deployment and operating costs, educational costs (e.g., specialized languages, training, or access to distributed systems and blockchain experts), and ease of partner onboarding and offboarding.

Unsurprisingly, given that the interviewees were selected for their interest in public cloud technologies, none of the respondents considered the ability to run a blockchain solution ``on premise'' a requirement; in fact, the overwhelming majority requested fully managed solutions, using terms such as ``\ac{SaaS}-style''. These requests were frequently coupled with concerns over operational and staffing complexity, with most interviewees acknowledging that, despite their interest, they were unprepared to staff, develop, or operate either public or private blockchain infrastructure at the time of interview. Another non-requirement we discovered was tokenization -- most interviewees agreed with the approach taken by Hyperledger Fabric and other permissioned solutions where nodes are treated as conventional enterprise infrastructure costs and there is no economic incentive desired or required for operation. Below we present two anecdotal but typical quotes gleaned from interviewees, and summarize the highest voted results from asking interviewees to select their top~5 requirements for distributed ledger technologies in table~\ref{tab:requirements}.

{\it\noindent{\bf CEO of a Leading Airline Alliance:} 
``To ensure appropriate and timely responses to market changes, businesses need to be highly agile, ensure connected experiences and tie cost to demand. We are a highly connected and complex industry and we succeed at delivering the best outcome to the travelers only if all partners are able to make decisions on a single, agreed upon version of the truth. Managing point solutions is expensive and the fixed costs are high and neither scalable nor agile. What we need is a highly scalable and agile multi-lateral agreement mechanism with a \ac{SaaS}-like model.''}

{\it\noindent{\bf CEO of a Leading Insurance Provider:} ``We need all the promises of Blockchain -- a single source of truth with each party controlling their own data -- but with the scale, cost advantages, and enterprise-grade feature set of a public cloud service.''}

Of the approximately 27\% of interviewees already engaged in any form of \ac{DLT} deployment (from prototyping through production attempts), the overwhelming majority reported a lack of success or significant impediments. This includes nearly 100\% churn among public-Ethereum-based trials and a striking 90\% abandonment rate for active PoCs, pilots, or other trials involving Hyperledger Fabric, with the remainder either incomplete at time of discussion or scoring poorly on likelihood of eventual implementation. The most frequently cited reasons for terminating a project were costs and complexity, with PoCs typically requiring 6-12 months and infrastructure and staffing or consulting costs that in many cases exceeded \$1M USD. Attempts to simulate partner onboarding registered the highest levels of complaints and failures, due to the additional costs, deployment, and connectivity burdens involved, which sometimes even led to scenarios in which one of the parties ran all the nodes in the blockchain network -- a setting that contradicts the original intention of a \ac{DLT}.

SOC2, GDPR, PCI, and other compliance programs that address regulation were typically cited as requirements, and interviewees also expressed \emph{de jeure} concerns: For example, the climate impact of \ac{PoW} solutions and its well-known energy consumption~\cite{sedlmeir2020energy} often failed to meet shareholder and customer expectations regarding a public corporation's environmental impact.

\section{A serverless DLT architecture}
\label{sec:serverless}
\begin{figure*}[!tbh]
    \centering
    \includegraphics[width=\textwidth, trim=0cm 0cm 0cm -1cm, clip]{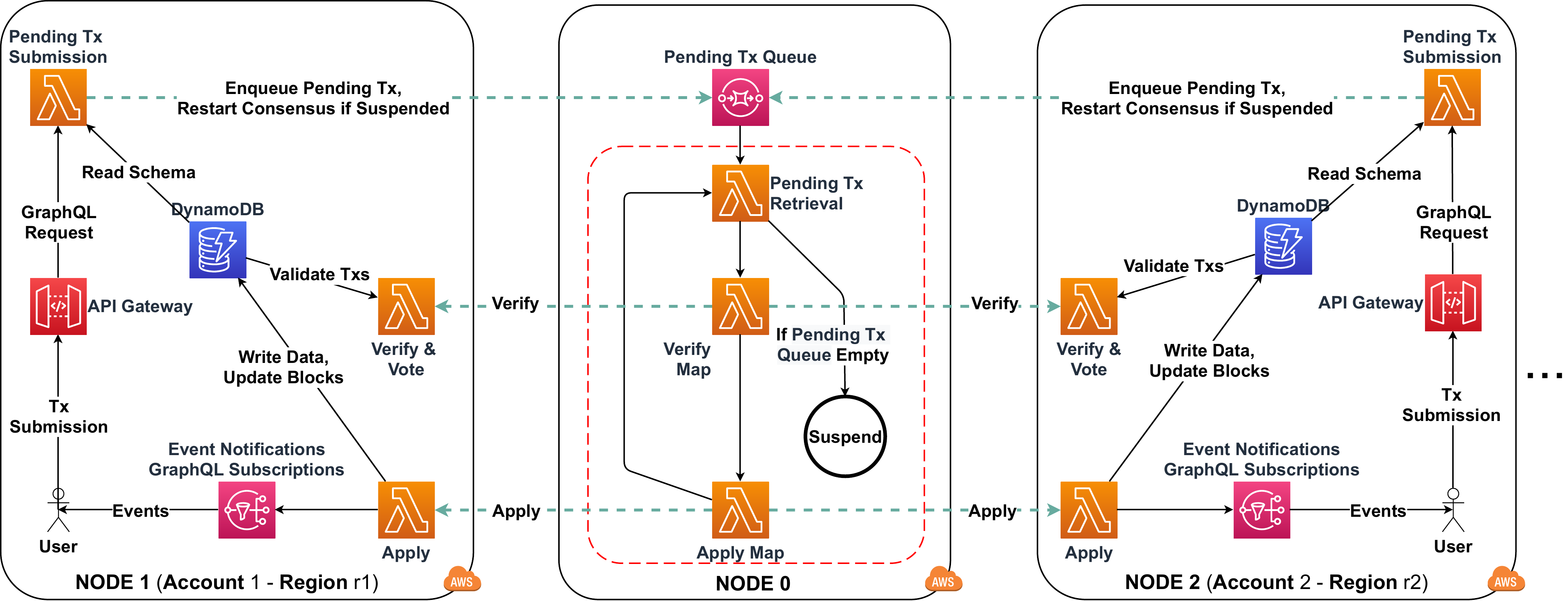}
    \caption{Components and transaction lifecycle of a serverless blockchain.}
    \label{fig:architecture}
\end{figure*}
Out of the box, Serverless cloud services share a key limitation with the earlier cloud technologies: A centralized, single-owner resource model. Moreover, these resources are mutable by the owner. Using them to construct a blockchain consensus algorithm and, thus, a multi-party, decentralized ledger requires algorithmic techniques that differ from both conventional consensus approaches and classic ``single party'' cloud application design. Figure~\ref{fig:architecture} illustrates the high-level architecture of the core of a serverless blockchain, analogous to the pending transaction ingestion, data replication and durability aspects, and consensus (``block minting'') elements of a conventional blockchain. Transactions follow a distributed two-phase commit lifecycle:

(1) An API Gateway receives the transaction as an HTTPS request, and uses a serverless function (\ac{AWS}~Lambda in our implementation) to add it to a (durable) shared pending transaction queue.
 Users can group their updates into ordered or unordered atomic transactions.
(2) A serverless choreography service reads pending transactions from the queue, combines them into a batch (``block'') by applying commutativity and associativity proofs that will enable non-deterministic parallelism, then orchestrates a two-phase commit among all nodes again using serverless functions. This ``leader'' function can be rotated amongst the nodes or operated in a separate account from them. (3) In the first (``verify'') phase, each node checks the syntactic and semantic validity of transactions;
the pending block is also written to durable storage.
(4) In the second (``apply'') phase, transactions are committed to world state and the block is marked committed.
In both phases, transactions within a block are processed in parallel, a key performance difference with prior approaches.
The ledger is a typical blockchain-style data structure in which each record contains a hash of its own content as well as a hash pointer to the previous block's content to provide tamper-evidence. Signed hashes from each of the nodes participating in the block's construction can be included to make the ledger a standalone ``proof of agreement'' that can be independently audited. Both ledger and world state are stored in a cloud-based NoSQL data store.

Conventional distributed application techniques can be incorporated into the algorithm above to enable individual nodes to fail (i.e., either verify or apply calls are not returned) and to re-synchronize them to the group, providing fault tolerance up to whatever degree (majority, Byzantine Attack-resistant majority, etc.) the chain's policy permits. 

A na\"ive implementation of the above sketch would require centralized trust: A nefarious orchestration, e.g., could ignore the actual votes from the first commit phase. To explore the construction of a decentralized approach, we define a simplified threat model we refer to as \emph{downward-only trust with identities}: Parties in the chain do not trust each other, but can reliably identify messages (either through the use of \ac{CSP} identity mechanisms or any form of key pairs). As with conventional blockchains, all parties trust their ``infrastructure'' -- e.g., assume that the CPU processors, data centers, and so forth faithfully execute the consensus algorithm as written. Network messages are assumed to be subject to loss and/or corruption by an adversary. Because our approach is cloud-based, denial of service (DOS) attacks are trivial to reject at the infrastructure level and do not appear in the consensus algorithm per se.

To counter threats, we rely on a combination of techniques to convert ``centralized'', single-owner computations into multi-party ones. Chief among these is \emph{verifiable immutability}: By rendering a resource immutable to everyone (including its creator) in a way that can be independently verified by others, trust in its content switches from an (untrusted) fellow participant in the chain to the (trusted) transitive closure of infrastructure. In conjunction with consensus, these techniques enable ``party-independent'' storage and compute along with the ability to verify both consensus correctness and enforce application-defined smart contract code reviews with effectively no performance overhead, i.e., $O(1)$ time relative to transaction submission and block construction.

\noindent\textbf{Code Storage:} Reference copies of code used to perform consensus and for user-provided smart contracts can be rendered immutable through the embargo features provided by all major \acs{CSP} blob storage services.

\noindent\textbf{Compute:} The versioning of cloud functions enables immutable execution, where the outcome is provably independent of the identity of both the owner and the caller. In addition, we rely on the ability of \acs{CSP}s to provide either the code or a hash for a function's content in a reliable way.

\noindent\textbf{Orchestrations:} We utilize \acs{CSP} immutability and/or versioning features to acquire a read-only copy of the orchestration that is guaranteed to be linked to an in-flight execution of same, and then prove its correctness by having verifiers vote on its veracity.

The serverless system that we developed includes a compiler capable of converting a JSON Schema-based representation of a data model into the multi-party, cloud-hosted deployment described above. Data integrity is handled as in prior approaches: hash chaining and signatures from all verifiers voting ``yes'' that include the block's id and hash protect against future attempts to corrupt, reorder, or repudiate transaction content and enable automated correctness proofs for materialized world state at any block height. By including software updates, metadata correctness proofs, smart contract code agreements, and data schema evolution as block entries, the trust model can be naturally extended to correctness proofs of the consensus algorithm itself (including software patches) and contract execution. Including transaction-submitted hashes and signatures extends this approach to protecting the individual content prior to submission.

\section{Qualitative evaluation}
\label{sec:qualitative_analysis}

We now discuss our approach to implementing a permissioned blockchain from a qualitative perspective, structured according to the business requirements collected in section~\ref{sec:requirements}. 

While our approach allows for individual nodes to be placed on the owner's preferred \acp{CSP}, conventional (``server-based'') blockchains also permit operating an individual node outside of any cloud, for example in a self-hosted data center. Recreating the fault tolerance inherent in our approach, would of course entail additional costs in that model to create multi-region data centers with decorrelated fault models. In choosing a cloud-native implementation, our approach restricts the set of providers to public cloud \acp{CSP}; our interview results indicated that companies are already reliant on or more \acp{CSP} for critical business processes and thus find this acceptable and in many cases preferable, as it simplifies deployment, management, hosting, and administration for them. The critical locus of trust for our survey respondents was with respect to other business parties participating in the chain, rather than whether the chain itself is hosted in the cloud or on premise, and they were comfortable trusting a provider such as \ac{AWS}, in much the same way they trust a company like Intel at the processor level. Consequently, our serverless distributed ledger approach can sufficiently address enterprises' \textbf{decentralization} needs.

Existing permissioned blockchains like Fabric or Quorum are known to be both CPU bound and to expose limited multi-CPU/multi-core parallelism, restricting their ability to scale elastically on demand~\cite{sedlmeir2021benchmarking}. Popular permissioned blockchains like Fabric and Quorum also employ databases, such as LevelDB or CouchDB, that have storage limits, unlike our approach's reliance on (the effectively limitless) cloud-hosted NoSQL database storage engines. According to several interviewees, LevelDB and CouchDB are uncommon in enterprise IT stacks, and having enterprise security teams authorize them can impose time-consuming analysis. Consequently, integrating a blockchain such as Hyperledger Fabric into a modern IT stack can require substantial infrastructure work, including networking, server allocation and maintenance, and long-term data storage considerations.
By contrast, our serverless blockchain approach is natively compatible with common cloud-based storage solutions, and the inherently multi-tenanted infrastructure and economies of scale enable our algorithm to exploit massively parallel data writing bandwidth to both NoSQL and blob storage services from within the compute layer. As a result, our ledgers, world state, and on-chain blob storage are all effectively unlimited -- \acp{CSP} simply grow their underlying physical data centers over time. Consequently, a serverless blockchain implementation addresses enterprise requirements regarding \textbf{elastic scaling on demand} and \textbf{unlimited storage} by design. 

A further benefit of using serverless technologies is that they natively incorporate \textbf{fault tolerance and high availability} into their implementations, relieving their owners of the responsibility of constructing and managing the associated scaling and monitoring infrastructure. This is also a cost optimization, as both the human and infrastructure costs of scaling and operating large fleets and then packing work into them is amortized across millions of users with heterogeneous loads. By contrast, a server-based \ac{DLT} must deploy multiple nodes (in the case of \ac{AWS}, e.g., typically three nodes in three \acp{AZ}) to achieve a 99.99\,\% availability \ac{SLA}, and scale vertically to peak load
requirements. Furthermore, while server-based nodes can crash, particularly under high load, the intrinsic fault tolerance of serverless computing methods admits to a distributed ledger design in which a single transaction or block might fail, but the system as a whole remains resilient, particularly as each resource (serverless function, orchestration, storage unit, etc.) offers fault tolerance independent of other components. By contrast, a node in a server-based blockchain implementation generally fails as a unit.

By design, a serverless distributed ledger addresses \textbf{cloud deployment} requirements, and features inherent to serverless resources also improve the \textbf{ease of deployment}: Measuring time-to-market objectively is a difficult exercise, but qualitatively a serverless approach is far simpler than one that exposes the details of server-based networking and infrastructure. In concrete terms, a serverless model allows not only the seamless integration with cloud-based services and legacy systems but also the reuse of well established building blocks, in particular, \ac{CSP} key distribution, \ac{IAM}, and production-grade security. In our approach, a multi-party, multi-region, multi-CSP production solution can be constructed and deployed from a data model in under 10 minutes, even up to hundreds of participants; similar approaches for highly available server-based blockchains, even when performed by highly experienced teams operating with large personnel and hardware budgets, would typically be in the range of weeks to multiple quarters, based on feedback from the interviews described in section~\ref{sec:requirements}. Our approach also supports fully managed (aka ``SaaS'') deployments, in which the accounts and resources associated with a given node are constructed by our system on behalf of that participant. 

Extant literature has already demonstrated that permissioned blockchains like Fabric and Quorum generally exhibit \textbf{low latency and fast finality}~\cite{sedlmeir2021benchmarking} on the order of several hundreds of milliseconds to a few seconds~\cite{sedlmeir2021benchmarking} and hence significantly improve on their public blockchain counterparts. While this is already suitable to address many enterprises' requirements, a serverless approach can further improve on these outcomes through the use of massively parallel computation and -- at least for intra-datacenter applications -- access to \ac{CSP} dark fiber. Moreover, permissioned blockchains' \textbf{energy consumption} is orders of magnitude below that of \ac{PoW}-cryptocurrencies, determined by the number of nodes as this reflects the degree of redundancy for the operation and storage of transactions~\cite{sedlmeir2020energy}. Our approach improves further on this result by collapsing energy consumption and costs to be linear in transaction processing, rather than a function of continuously operated peak capacity; this is made possible by employing a highly multi-tenanted substrate that can effectively share compute, storage, and network capacity across many users with spatially and temporally decorrelated workloads.

Privacy and \textbf{access control}, especially the ability of a transaction's submitter to subset the viewers or updaters of its content amongst chain participants, is a critical enterprise feature. For example, Quorum and Hyperledger Fabric support private transactions that are only stored and executed in non-obfuscated form by the intended recipients~\cite{androulaki2018hyperledger}. 
We have extended our verifiable immutability approach to include enforcement of policies, using it to create \acp{ACL} on all fields, regardless of size or data type. \acp{ACL} are themselves stored in the ledger, enabling full auditing and lineage tracking for permission-based metadata in the same way that the underlying data itself is managed and queried. While this approach yields essentially the same functionality as private transactions offered by some existing permissioned blockchains, it considerably simplifies the ease of deployment, for example, compared to Quorum where the additional setup of a software-based enclave is necessary, or Fabric, where access control lists need to be specified individually for every smart contract at the time of deployment. 

To further substantiate our qualitative arguments, we also selected a subset of~50 companies previously interviewed and presented our vision of a serverless distributed ledger. 45~of those indicated that the approach was ‘likely’ or ‘very likely’ to meet their needs, and 10 companies have already piloted or deployed a commercialization of this approach, half of which used it to replace Ethereum or Hyperledger.

\section{Quantitative evaluation}
\label{sec:quantitative_analysis}

To examine the \textbf{performance} effect of multi-machine parallelism and access to massive data transfer parallelism available in the public cloud, we compared our approach to two permissioned blockchains, Fabric and Quorum, as these generally exhibited the best performance among several permissioned blockchains in a performance analysis~\cite{sedlmeir2021benchmarking}. For our benchmarks, we leveraged the distributed ledger performance scan (DLPS), a standardized tool for determining maximum throughput, latency, and resource metrics~\cite{sedlmeir2021benchmarking}.
We set up a Fabric network with 8~peers and 4~orderers, and a Quorum network of~8 nodes to compare with an 8-``node'' serverless blockchain. We investigated different choices of hardware in \ac{AWS} for the server-based blockchains, and chose a simple transaction payload in all cases (writing a single key-value pair). All benchmarks were conducted with default user account settings in \ac{AWS}, with no limit increases. We tested single-datacenter deployments, cross-European deployments with two datacenters in Frankfurt and Dublin, and an intercontinental setup with four datacenters in Singapore, Sao Paolo, Frankfurt, and Virginia to explore geo-related latency sensitivities. Even for very expensive hardware (16 vCPUs per node), the maximum throughput of Fabric and Quorum did not exceed 4,000\,tx/s (compare also~\cite{baliga2018quorum,androulaki2018hyperledger,thakkar2020scaling}), with latencies of around~1-2~seconds. Significantly, \emph{a request rate in excess of the maximum throughput frequently leads to the crash of at least one node within less than a minute} (see also the discussion of local fault tolerance in section~\ref{sec:qualitative_analysis}), requiring manual intervention to recover. By contrast, the serverless blockchain achieved a maximum ingress rate of more than 8,000\,tx/s in all scenarios (up to 75,000\,tx/s in the single datacenter scenario), and remained resilient well beyond this rate. 
Our initial prototype, without commit-time parallelism, was limited to 200~tx/s due to its use of an off-the-shelf cloud orchestration service. Preliminary results from rewriting our consensus in the form of cloud functions indicate that we can effectively parallelize thousands of world state updates per block, effectively exploiting the massively parallel data-planes available in cloud-based NoSQL data stores to match ingestion rates. We consequently expect that we can reach a commit throughput of several thousands of transactions per second in an optimized version, and more than 10,000 transactions per second with customized \ac{CSP} account settings.

For a server-based blockchain, the operating \textbf{cost} per transaction related to infrastructure is straightforward to compute: Assuming the same hardware for all nodes, the costs per second are simply the costs for all servers per second. 
Low throughput, thus, means that costs per transaction are high. When attempted throughput approaches maximum throughput, the costs for a server-based blockchain can become very low; however, our results (see above) suggest these systems become increasingly unreliable when actual loads near maximum capacity.
In contrast, owing to the multiple components that are invoked and billed separately during the lifecycle of a transaction, the cost structure for a serverless blockchain is considerably more complex.
In general, the cost structure in our approach involved both a fixed overhead per transaction and both fixed and variable per-block overhead.
For small request frequencies, block sizes will typically have a single transaction ($n$~=~1), and experimentally, we determined a cost of~\$\,0.0001; at the maximum packing size $n$~$\approx$~900, of our tested implementation, amortizing block costs reduced this to \$\,0.00001 on a per transaction basis. Unsurprisingly, compute costs (cloud function invocations) dominated our cost structure in this experimental setup, yet larger payload sizes could alter that in favor of data transfer or storage costs.
An interesting characteristic of our approach is that it allows clients to express their latency sensitivity: While a transaction that needs low latency cannot be more expensive than the costs for $n$~=~1, specifying a high latency bound could enable lower transaction prices in the presence of infrequent arrival rates by allowing larger blocks to be minted. 

\begin{figure}[!tb]
    \centering
    \includegraphics[width=0.8\linewidth]{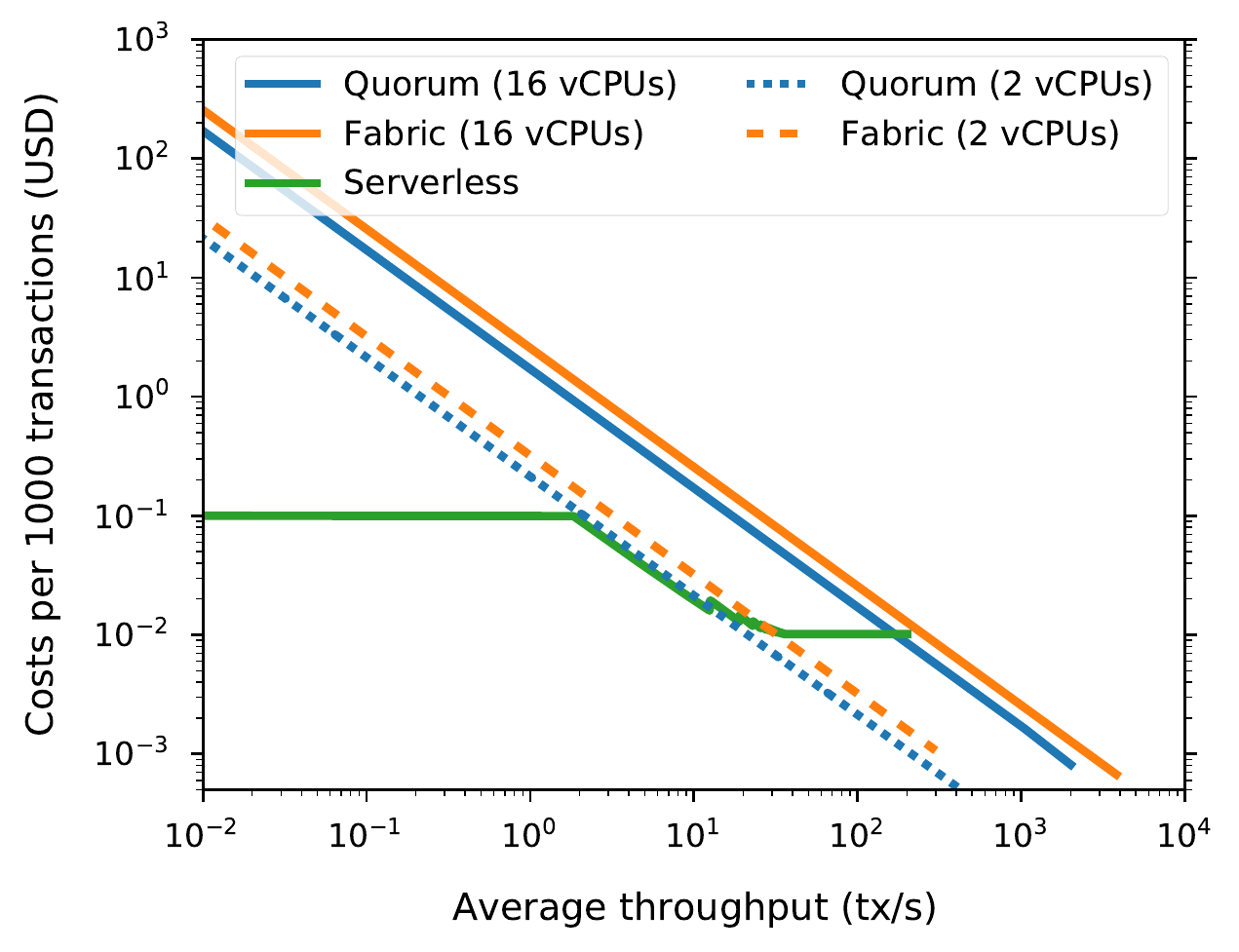}
    \caption{Comparison of per-transaction costs for serverful and serverless distributed ledgers.}
    \label{fig:cost_comparison}
\end{figure}

Figure~\ref{fig:cost_comparison} compares per-transaction costs for Fabric, Quorum and our serverless solution based on average throughput. The costs of our serverless implementation have caps due to the minimum and maximum batch size and interpolate cost at intermediate batch sizes. For infrequent arrival rates, serverless significantly outperforms server-based solutions; conversely, server-based solutions shine when they are operated just below their point of failure. Many corporate application workloads are not constant, of course, and as volatility increases, the ratio of maximum throughput to average throughput increases. For server-based blockchains, this requires more expensive hardware (in the form of vertical scaling), increasing per-transaction costs, whereas serverless solutions approximate costs that are linear in the number of transactions processed regardless of scale or volatility.
Finally, for this comparison we did \emph{not} create the 3-way redundancy required to achieve an approximately equivalent level of fault tolerance on Fabric or Quorum; multiplying the costs of those systems by 3X would yield a comparable outcome in this regard, strongly favoring the serverless approach.
\section{Conclusions and future research}
\label{sec:conclusion}
In this paper, we collect enterprise requirements for blockchains to enable cross-organization data exchange and propose an approach that combines many of the decentralization benefits of conventional distributed ledger approaches with the advantages of multi-tenanted but centralized cloud services. While blockchains may be employed for a wide
variety of purposes, our approach aligns with the needs of business users attempting to construct a ``single source of truth'' among untrusted business parties. Our contributions include exploration of blended approaches that lie neither in centralized nor conventional (``server-based'') decentralized algorithms, and which are capable of exploiting massive multi-machine parallelism to overcome scaling and smart contract processing limitations inherent in single-box approaches, while still exhibiting useful decentralized outcomes such as isolation and consistent data replication among nodes. Benchmarking of two popular permissioned blockchains, Fabric and Quorum, against our serverless implementation in terms of throughput and costs indicates that our current implementation already improves on multiple performance aspects -- including transaction ingress rates well in excess of those achievable through conventional means, while further optimizations promise to outperform existing permissioned blockchains through readily exploited avenues, such as decoupling of transaction content copying from consensus. Future work will also focus on establishing lower bounds for blockchains in which compute, storage, and network capacity are effectively unbounded, such as highly parallelizeable associativity and commutativity proofs, and a quantitative study of smart contract performance comparisons to conventional approaches. We also aim to rigorously evaluate whether the projects that leverage our serverless distributed ledger will have a higher success rate than what we found for existing permissioned blockchains in our interview study.

Our initial results were produced on \ac{AWS}, and some of the features on which we relied, such as function versioning, are not fully implemented on other providers, requiring additional or alternative approaches. More interesting as a research avenue is cross-cloud fault tolerance, in which consensus can span \acp{CSP} and survive temporary outages in much the same way that the existing system can survive regional outages \emph{within} a \ac{CSP}. We hypothesize that selective use of conventional consensus algorithms across clouds could be applied in such a way as to minimize the performance impact while offering enhanced threat models and availability guarantees and hope to explore such patterns in future work.7

\bibliographystyle{ieeetr}
\bibliography{references}

\begin{thebibliography}{10}

\bibitem{lamport1982byzantine}
L.~Lamport, R.~Shostak, and M.~Pease, ``{The Byzantine Generals Problem},''
  {\em ACM Transactions on Programming Languages and Systems}, vol.~4, no.~3,
  pp.~382--401, 1982.

\bibitem{Castro:1999:pbft}
M.~Castro, B.~Liskov, {\em et~al.}, ``{Practical Byzantine Fault Tolerance},''
  in {\em Proceedings of the Third Symposium on Operating Systems Design and
  Implementation}, pp.~173--186, 1999.

\bibitem{ailijiang2016consensus}
A.~Ailijiang, A.~Charapko, and M.~Demirbas, ``{Consensus in the Cloud: Paxos
  Systems Demystified},'' in {\em 25th International Conference on Computer
  Communication and Networks}, 2016.

\bibitem{Nakamoto:2008:Bitcoin}
S.~Nakamoto, ``{Bitcoin: A Peer-to-peer Electronic Cash System},'' 2008.

\bibitem{Buterin:2014:Ethereum}
V.~Buterin {\em et~al.}, ``{A Next-Generation Smart Contract and Decentralized
  Application Platform},'' 2014.

\bibitem{beck2018governance}
R.~Beck, C.~M{\"u}ller-Bloch, and J.~L. King, ``{Governance in the Blockchain
  Economy: A Framework and Research Agenda},'' {\em Journal of the Association
  for Information Systems}, vol.~19, no.~10, pp.~1020--1034, 2018.

\bibitem{fridgen2018cross}
G.~Fridgen, S.~Radszuwill, N.~Urbach, and L.~Utz, ``{Cross-organizational
  Workflow Management using Blockchain Technology -- Towards Applicability,
  Auditability, and Automation},'' in {\em Proceedings of the 51st Hawaii
  International Conference on System Sciences}, 2018.

\bibitem{luu2015demystifying}
L.~Luu, J.~Teutsch, R.~Kulkarni, and P.~Saxena, ``{Demystifying Incentives in
  the Consensus Computer},'' in {\em Proceedings of the 22nd ACM SIGSAC
  Conference on Computer and Communications Security}, pp.~706--719, 2015.

\bibitem{kannengiesser2020trade}
N.~Kannengie{\ss}er, S.~Lins, T.~Dehling, and A.~Sunyaev, ``{Trade-offs between
  Distributed Ledger Technology Characteristics},'' {\em ACM Computing
  Surveys}, vol.~53, no.~2, 2020.

\bibitem{zhang2019security}
R.~Zhang, R.~Xue, and L.~Liu, ``{Security and Privacy on Blockchain},'' {\em
  ACM Computing Surveys}, vol.~52, no.~3, 2019.

\bibitem{miehle2019partchain}
D.~Miehle, D.~Henze, A.~Seitz, A.~Luckow, and B.~Bruegge, ``{PartChain: A
  Decentralized Traceability Application for Multi-Tier Supply Chain Networks
  in the Automotive Industry},'' in {\em International Conference on
  Decentralized Applications and Infrastructures}, pp.~140--145, 2019.

\bibitem{androulaki2018hyperledger}
E.~Androulaki, A.~Barger, V.~Bortnikov, C.~Cachin, K.~Christidis, A.~De~Caro,
  {\em et~al.}, ``{Hyperledger Fabric: A Distributed Operating System for
  Permissioned Blockchains},'' in {\em ACM Proceedings of the Thirteenth
  EuroSys Conference}, 2018.

\bibitem{barr_2019}
J.~Barr, ``{Amazon Prime Day 2019 – Powered by AWS},'' 2019.

\bibitem{castro2019rise}
P.~Castro, V.~Ishakian, V.~Muthusamy, and A.~Slominski, ``{The Rise of
  Serverless Computing},'' {\em Communications of the ACM}, vol.~62, no.~12,
  pp.~44--54, 2019.

\bibitem{jonas2019cloud}
E.~Jonas, J.~Schleier-Smith, V.~Sreekanti, C.-C. Tsai, A.~Khandelwal, Q.~Pu,
  V.~Shankar, J.~Carreira, K.~Krauth, N.~Yadwadkar, J.~E. Gonzalez, R.~A. Popa,
  I.~Stoica, and D.~A. Patterson, ``{Cloud Programming Simplified: A Berkeley
  View on Serverless Computing},'' 2019.

\bibitem{schleier-smith2021serverless}
J.~Schleier-Smith, V.~Sreekanti, A.~Khandelwal, J.~Carreira, N.~J. Yadwadkar,
  R.~A. Popa, J.~E. Gonzalez, I.~Stoica, and D.~A. Patterson, ``{What
  Serverless Computing Is and Should Become: The Next Phase of Cloud
  Computing},'' {\em Communications of the ACM}, vol.~64, no.~5, pp.~76--84,
  2021.

\bibitem{451ResearchUtilization}
{451 Research}, ``{The Carbon Reduction Opportunity of Moving to Amazon Web
  Services},'' 2019.

\bibitem{zavolokina2020management}
L.~Zavolokina, R.~Ziolkowski, I.~Bauer, and G.~Schwabe, ``{Management,
  Governance and Value Creation in a Blockchain Consortium},'' {\em MIS
  Quarterly Executive}, vol.~19, no.~1, pp.~1--17, 2020.

\bibitem{lacity2018addressing}
M.~C. Lacity, ``{Addressing Key Challenges to Making Enterprise Blockchain
  Applications a Reality},'' {\em MIS Quarterly Executive}, vol.~17, no.~3,
  pp.~201--222, 2018.

\bibitem{hastig2020blockchain}
G.~M. Hastig and M.~S. Sodhi, ``{Blockchain for Supply Chain Traceability:
  Business Requirements and Critical Success Factors},'' {\em Production and
  Operations Management}, vol.~29, no.~4, pp.~935--954, 2020.

\bibitem{chatterjee2019challenges}
A.~Chatterjee, M.~Parmar, and Y.~Pitroda, ``{Production Challenges of
  Distributed Ledger Technology (DLT) based Enterprise Applications},'' in {\em
  International Symposium on Systems Engineering}, 2019.

\bibitem{oh2019serverless}
B.~Oh and D.~Kim, ``{Serverless-Enabled Permissioned Blockchain for Elastic
  Transaction Processing},'' in {\em Proceedings of the 20th International
  Middleware Conference Demos and Posters}, pp.~9--–10, 2019.

\bibitem{ghaemi2020chainfaas}
S.~Ghaemi, H.~Khazaei, and P.~Musilek, ``{ChainFaaS: An Open Blockchain-based
  Serverless Platform},'' {\em IEEE Access}, vol.~8, pp.~131760--131778, 2020.

\bibitem{kaplunovich2019scalability}
A.~Kaplunovich, K.~P. Joshi, and Y.~Yesha, ``{Scalability Analysis of
  Blockchain on a Serverless Cloud},'' in {\em IEEE International Conference on
  Big Data}, pp.~4214--4222, 2019.

\bibitem{antonopoulos2021azure}
P.~Antonopoulos, R.~Kaushik, H.~Kodavalla, S.~R. Aceves, R.~Wong, J.~Anderson,
  and J.~Szymaszek, ``{SQL Ledger: Cryptographically Verifiable Data in Azure
  SQL Database},'' in {\em Proceedings of SIGMOD}, ACM, 2021.

\bibitem{sedlmeir2020energy}
J.~Sedlmeir, H.~U. Buhl, G.~Fridgen, and R.~Keller, ``{The Energy Consumption
  of Blockchain Technology: Beyond Myth},'' {\em Business \& Information
  Systems Engineering}, vol.~62, no.~6, pp.~599--608, 2020.

\bibitem{sedlmeir2021benchmarking}
J.~Sedlmeir, P.~Ross, A.~Luckow, J.~Lockl, D.~Miehle, and G.~Fridgen, ``{The
  DLPS: A Framework for Benchmarking Blockchains},'' in {\em Proceedings of the
  54th Hawaii International Conference on System Sciences}, pp.~6855--6864,
  2021.

\bibitem{baliga2018quorum}
A.~Baliga, I.~Subhod, P.~Kamat, and S.~Chatterjee, ``{Performance Evaluation of
  the {Quorum} Blockchain Platform},'' 2018.

\bibitem{thakkar2020scaling}
P.~Thakkar and S.~Nathan, ``{Scaling Hyperledger Fabric Using Pipelined
  Execution and Sparse Peers},'' 2020.

\end{thebibliography}

\end{document}